\documentclass[preprint2]{aastex}

\newcommand{\myemail}{beilicke@physics.wustl.edu}
\slugcomment{The Astrophysical Journal}

\shorttitle{VERITAS GC Observations}
\shortauthors{VERITAS Collaboration}

\begin{document}

\title{Very-high energy observations of the Galactic center region by 
VERITAS in 2010--2012}

\author{
A.~Archer\altaffilmark{1},
A.~Barnacka\altaffilmark{2},
M.~Beilicke\altaffilmark{1,*},
W.~Benbow\altaffilmark{3},
K.~Berger\altaffilmark{4},
R.~Bird\altaffilmark{5},
J.~Biteau\altaffilmark{6},
J.~H.~Buckley\altaffilmark{1},
V.~Bugaev\altaffilmark{1},
K.~Byrum\altaffilmark{7},
J.~V~Cardenzana\altaffilmark{8},
M.~Cerruti\altaffilmark{3},
W.~Chen\altaffilmark{1},
X.~Chen\altaffilmark{9,10},
L.~Ciupik\altaffilmark{11},
M.~P.~Connolly\altaffilmark{12},
W.~Cui\altaffilmark{13},
H.~J.~Dickinson\altaffilmark{8},
J.~Dumm\altaffilmark{14},
J.~D.~Eisch\altaffilmark{8},
A.~Falcone\altaffilmark{15},
S.~Federici\altaffilmark{10,9},
Q.~Feng\altaffilmark{13},
J.~P.~Finley\altaffilmark{13},
H.~Fleischhack\altaffilmark{10},
L.~Fortson\altaffilmark{14},
A.~Furniss\altaffilmark{6},
N.~Galante\altaffilmark{3},
S.~Griffin\altaffilmark{16},
S.~T.~Griffiths\altaffilmark{17},
J.~Grube\altaffilmark{11},
G.~Gyuk\altaffilmark{11},
N.~H{\aa}kansson\altaffilmark{9},
D.~Hanna\altaffilmark{16},
J.~Holder\altaffilmark{4},
G.~Hughes\altaffilmark{10},
C.~A.~Johnson\altaffilmark{6},
P.~Kaaret\altaffilmark{17},
P.~Kar\altaffilmark{18},
M.~Kertzman\altaffilmark{19},
Y.~Khassen\altaffilmark{5},
D.~Kieda\altaffilmark{18},
H.~Krawczynski\altaffilmark{1},
S.~Kumar\altaffilmark{4},
M.~J.~Lang\altaffilmark{12},
G.~Maier\altaffilmark{10},
S.~McArthur\altaffilmark{20},
A.~McCann\altaffilmark{21},
K.~Meagher\altaffilmark{22},
P.~Moriarty\altaffilmark{23,12},
R.~Mukherjee\altaffilmark{24},
D.~Nieto\altaffilmark{25},
A.~O'Faol\'{a}in de Bhr\'{o}ithe\altaffilmark{10},
R.~A.~Ong\altaffilmark{26},
A.~N.~Otte\altaffilmark{22},
N.~Park\altaffilmark{20},
J.~S.~Perkins\altaffilmark{27},
M.~Pohl\altaffilmark{9,10},
A.~Popkow\altaffilmark{26},
H.~Prokoph\altaffilmark{10},
E.~Pueschel\altaffilmark{5},
J.~Quinn\altaffilmark{5},
K.~Ragan\altaffilmark{16},
J.~Rajotte\altaffilmark{16},
L.~C.~Reyes\altaffilmark{28},
P.~T.~Reynolds\altaffilmark{29},
G.~T.~Richards\altaffilmark{22},
E.~Roache\altaffilmark{3},
G.~H.~Sembroski\altaffilmark{13},
K.~Shahinyan\altaffilmark{14},
A.~W.~Smith\altaffilmark{18},
D.~Staszak\altaffilmark{16},
I.~Telezhinsky\altaffilmark{9,10},
J.~V.~Tucci\altaffilmark{13},
J.~Tyler\altaffilmark{16},
A.~Varlotta\altaffilmark{13},
S.~Vincent\altaffilmark{10},
S.~P.~Wakely\altaffilmark{20},
A.~Weinstein\altaffilmark{8},
R.~Welsing\altaffilmark{10},
A.~Wilhelm\altaffilmark{9,10},
D.~A.~Williams\altaffilmark{6},
A.~Zajczyk\altaffilmark{1},
B.~Zitzer\altaffilmark{7}
}

\altaffiltext{1}{Department of Physics, Washington University, St. Louis, MO 63130, USA}
\altaffiltext{2}{Harvard-Smithsonian Center for Astrophysics, 60 Garden Street, Cambridge, MA 02138, USA}
\altaffiltext{3}{Fred Lawrence Whipple Observatory, Harvard-Smithsonian Center for Astrophysics, Amado, AZ 85645, USA}
\altaffiltext{4}{Department of Physics and Astronomy and the Bartol Research Institute, University of Delaware, Newark, DE 19716, USA}
\altaffiltext{5}{School of Physics, University College Dublin, Belfield, Dublin 4, Ireland}
\altaffiltext{6}{Santa Cruz Institute for Particle Physics and Department of Physics, University of California, Santa Cruz, CA 95064, USA}
\altaffiltext{7}{Argonne National Laboratory, 9700 S. Cass Avenue, Argonne, IL 60439, USA}
\altaffiltext{8}{Department of Physics and Astronomy, Iowa State University, Ames, IA 50011, USA}
\altaffiltext{9}{Institute of Physics and Astronomy, University of Potsdam, 14476 Potsdam-Golm, Germany}
\altaffiltext{10}{DESY, Platanenallee 6, 15738 Zeuthen, Germany}
\altaffiltext{11}{Astronomy Department, Adler Planetarium and Astronomy Museum, Chicago, IL 60605, USA}
\altaffiltext{12}{School of Physics, National University of Ireland Galway, University Road, Galway, Ireland}
\altaffiltext{13}{Department of Physics and Astronomy, Purdue University, West Lafayette, IN 47907, USA}
\altaffiltext{14}{School of Physics and Astronomy, University of Minnesota, Minneapolis, MN 55455, USA}
\altaffiltext{15}{Department of Astronomy and Astrophysics, 525 Davey Lab, Pennsylvania State University, University Park, PA 16802, USA}
\altaffiltext{16}{Physics Department, McGill University, Montreal, QC H3A 2T8, Canada}
\altaffiltext{17}{Department of Physics and Astronomy, University of Iowa, Van Allen Hall, Iowa City, IA 52242, USA}
\altaffiltext{18}{Department of Physics and Astronomy, University of Utah, Salt Lake City, UT 84112, USA}
\altaffiltext{19}{Department of Physics and Astronomy, DePauw University, Greencastle, IN 46135-0037, USA}
\altaffiltext{20}{Enrico Fermi Institute, University of Chicago, Chicago, IL 60637, USA}
\altaffiltext{21}{Kavli Institute for Cosmological Physics, University of Chicago, Chicago, IL 60637, USA}
\altaffiltext{22}{School of Physics and Center for Relativistic Astrophysics, Georgia Institute of Technology, 837 State Street NW, Atlanta, GA 30332-0430}
\altaffiltext{23}{Department of Life and Physical Sciences, Galway-Mayo Institute of Technology, Dublin Road, Galway, Ireland}
\altaffiltext{24}{Department of Physics and Astronomy, Barnard College, Columbia University, NY 10027, USA}
\altaffiltext{25}{Physics Department, Columbia University, New York, NY 10027, USA}
\altaffiltext{26}{Department of Physics and Astronomy, University of California, Los Angeles, CA 90095, USA}
\altaffiltext{27}{N.A.S.A./Goddard Space-Flight Center, Code 661, Greenbelt, MD 20771, USA}
\altaffiltext{28}{Physics Department, California Polytechnic State University, San Luis Obispo, CA 94307, USA}
\altaffiltext{29}{Department of Applied Physics and Instrumentation, Cork Institute of Technology, Bishopstown, Cork, Ireland}
\altaffiltext{*}{\myemail}

%% ############################################################
%% ############ Abstract
%% ############################################################
\begin{abstract}

The Galactic center is an interesting region for high-energy ($0.1-100 
\, \rm{GeV}$) and very-high-energy ($E > 100 \, \rm{GeV}$) $\gamma$-ray 
observations. Potential sources of GeV/TeV $\gamma$-ray emission have 
been suggested, e.g., the accretion of matter onto the supermassive 
black hole, cosmic rays from a nearby supernova remnant (e.g. 
Sgr\,A~East), particle acceleration in a plerion, or the annihilation of 
dark matter particles. The Galactic center has been detected by EGRET 
and by {\it Fermi}/LAT in the MeV/GeV energy band. At TeV energies, the 
Galactic center was detected with moderate significance by the CANGAROO 
and Whipple $10 \, \rm{m}$ telescopes and with high significance by 
H.E.S.S., MAGIC, and VERITAS. We present the results from three years of 
VERITAS observations conducted at large zenith angles resulting in a 
detection of the Galactic center on the level of $18$ standard 
deviations at energies above $\sim$$2.5 \, \rm{TeV}$. The energy 
spectrum is derived and is found to be compatible with hadronic, 
leptonic and hybrid emission models discussed in the literature. Future, 
more detailed measurements of the high-energy cutoff and better 
constraints on the high-energy flux variability will help to refine 
and/or disentangle the individual models.

\end{abstract}

\keywords{gamma-rays --- Galactic center --- black hole --- non-thermal 
--- VERITAS --- sources: VER\,J1745-290}


%% ############################################################
%% ############################################################
%% ############ Introduction
%% ############################################################
%% ############################################################
\section{Introduction}
\label{sec:Introduction}

The strong radio source Sgr\,A* located in the center of our galaxy is 
believed to coincide with a $4 \times 10^{6} \, M_{\odot}$ black hole. 
While molecular clouds and dust hide the view towards the Galactic 
center at optical wavelengths, transient X-ray events with a $2-10 \, 
\rm{keV}$ energy output up to $10^{35} \, \rm{ergs}/s$ are observed from 
Sgr\,A* on a regular basis \citep{Swift_X-rayFlares, Chandra2012, 
NuStar2012}, as well as transient events at MeV/GeV energies 
\citep{Fermi_GC_Burst}. Flares from X-ray binaries located in the 
Galactic center region can reach luminosities up to $10^{37} \, 
\rm{ergs}/s$ \citep{Muno2005, Porquet2005, Sakano2005, Wijnands2006, 
Degenaar2012}. Various astrophysical sources located close to the 
Galactic center are potentially capable of accelerating particles to 
multi-TeV energies, such as the supernova remnant Sgr\,A~East or the 
pulsar wind nebula (PWN) G\,359.95-0.04 \citep{GC_Plerion}.

A recently discovered gaseous object, G\,2, heading towards the 
immediate vicinity of the Galactic center \citep{BH_eats_MC} is expected 
to start merging into the black hole accretion stream some time in 
2013-2014. This potential merger is a once-in-a-lifetime event that will 
allow observers to test magneto-hydrodynamical accretion models and 
their potential link to emission at the highest energies. Simulations 
show that the expected change in accretion and emission strongly depend 
on the origin/properties of the object \citep{G2_Simulations, 
G2_Simulations2, G2_Simulations3} which have not yet been constrained 
well enough for accurate predictions. The merging process can 
potentially last for several decades and represents strong motivation 
for establishing a baseline for the $\gamma$-ray emission (as presented 
here) as well as for ongoing long-term monitoring of this region.

% JB - One might argue that the ordinary matter does not bind the DM halo, 
% but the other way around.

Observations of the Galactic center region also provide an avenue for 
dark-matter detection \citep{HESS_GC_DM}.  Cold dark matter is widely 
viewed to be an essential component of the Universe in our current 
standard cosmological model. A $100 \, \rm{GeV}$ to TeV scale thermal 
relic with weak-scale interactions (or weakly interacting massive 
particle, WIMP) could provide the cold dark matter required to explain 
the observed structure in the Universe as well as the matter density 
derived from cosmic microwave background measurements. However, to 
effectively search for a dark matter annihilation signal in the GeV/TeV 
regime, it is necessary to first understand the distribution, angular 
extent, and energy spectrum of the astrophysical sources near the 
Galactic center.

Several astrophysical sources located in the vicinity of the Galactic 
center can potentially emit $\gamma$-rays at MeV/GeV/TeV energies. 
Definite associations, on the other hand, are hampered by the limited 
angular resolution of instruments in these wave bands, ranging from 
$\simeq 0.1 \, \deg$ at TeV energies to several degrees in the MeV 
regime. The EGRET MeV/GeV $\gamma$-ray source 3EG\,J1746-2851 is 
spatially coincident with the Galactic center \citep{Egret_GC}. More 
than one MeV/GeV source were resolved by the {\it Fermi}/LAT instrument 
($20 \, \rm{MeV} < E \lesssim 100 \, \rm{GeV}$) in the inner $\sim$$3 
\deg$ region around the Galactic center \citep{Fermi_FirstCatalog, 
Fermi_SecondCatalog}, with the strongest source being spatially 
coincident with the Galactic center (see sky map in 
Sec.~\ref{sec:GC_VERITAS}). Uncertainties in the diffuse Galactic 
background models and the limited angular resolution of the {\it 
Fermi}/LAT limits the ability to study the morphologies of these MeV/GeV 
sources in great detail.

At GeV/TeV energies a detection of a source coincident with the position 
of the Galactic center was first reported by the CANGAROO\,II 
collaboration which operated a ground-based $\gamma$-ray telescope in 
the southern hemisphere. The CANGAROO collaboration reported a steep 
energy spectrum $\rm{d}N/\rm{d}E \propto E^{-4.6}$ ($250 \, \rm{GeV} < E 
\lesssim 2.5 \, \rm{TeV}$) with an integral flux above $250 \, \rm{GeV}$ 
at the level of $10\%$ of the Crab Nebula flux \citep{CANGAROO_GC}. 
Shortly after, evidence for emission above $2.8 \, \rm{TeV}$ from the 
Galactic center at the level of $3.7$ standard deviations (s.\,d.) was 
reported from 1995-2003 large zenith angle (LZA) observations ($2.8 \, 
\rm{TeV} < E \lesssim 10 \, \rm{TeV}$) with the Whipple $10 \, \rm{m}$ 
$\gamma$-ray telescope \citep{Whipple_GC}. (see Sec.~\ref{sec:LZA} for 
an explanation of LZA observations).

Observations by H.E.S.S.~in 2004-2006 confirmed the Galactic center as a 
GeV/TeV $\gamma$-ray source in the energy range of $100 \, \rm{GeV}$ to 
several tens of TeV \citep{HESS_SgrA}. The measured energy spectrum is 
described by a power law $\rm{d}N/\rm{d}E \propto E^{-2.1}$ with a 
cut-off at $\sim$$15 \, \rm{TeV}$ \citep{HESS_GC_Spectrum}. No evidence 
for variability was found in the H.E.S.S. or Whipple data over a time 
span of more than ten years. The difference between the energy spectrum 
measured by CANGAROO compared to the spectra measured by the other 
ground-based GeV/TeV instruments could only be explained if the 
different instruments observed different astrophysical sources in 
different states of activity or if the CANGAROO results were affected by 
a measurement error (see e.g. \citet{CangarooSystematic}).

Using the high-precision pointing system of the H.E.S.S. telescopes 
(reducing the pointing uncertainty to $6 \arcsec$ per axis), the 
position of the supernova remnant Sgr\,A~East could be excluded as the 
source of the TeV $\gamma$-ray emission \citep{HESS_GC_Location}. A 
diffuse GeV/TeV $\gamma$-ray emission was identified after subtracting 
the point source located at the position of the Galactic center 
\citep{HESS_SgrA_Diffuse}. Its intensity profile is aligned along the 
Galactic plane (see sky map in Sec.~\ref{sec:GC_VERITAS}) and follows 
the structure of molecular clouds. The energy spectrum of the diffuse 
emission is described by a power law $\rm{d}N/\rm{d}E \propto E^{-2.3}$. 
It can be explained by an interaction of local cosmic rays (CRs) with 
the matter in the molecular clouds~-- indicating a harder spectrum and a 
higher flux of CRs in this inner region of the Galaxy as compared to the 
local CR spectrum ($\rm{d}N/\rm{d}E \propto E^{-2.7}$) observed at 
Earth. Recently, an additional, unresolved diffuse component of 
$\gamma$-ray emission was identified by the H.E.S.S.~collaboration along 
the extended Galactic plane \citep{HESS_UnresolvedDiffuse}. The MAGIC 
collaboration detected the Galactic center in 2004/05 observations 
performed at large zenith angles at the level of $7$ s.\,d.~($0.5 - 10 
\, \rm{TeV}$) \citep{MAGIC_GC}, confirming the energy spectrum measured 
by H.E.S.S.

VERITAS first reported a $> 10$ s.\,d.~detection of the Galactic center 
in 2010 LZA observations, covering an energy range of $2.5 \, \rm{TeV}$ 
to several tens of TeV \citep{Beilicke2011}. In this paper we report on 
the results of three years (2010-2012) of VERITAS observations of the 
Galactic center region at large zenith angles at energies above 
$\sim$$2.5 \, \rm{TeV}$. This paper focuses on the central TeV 
$\gamma$-ray source coincident with the Galactic center. The data were 
analyzed with the {\it displacement} method which substantially improves 
the angular resolution and sensitivity for data taken at large zenith 
angles (see Sec.~\ref{sec:LZA}). The VERITAS observations and results 
are discussed in Sec.~\ref{sec:GC_VERITAS}. A discussion, comparison to 
models and prospects of future GeV/TeV $\gamma$-ray observations of the 
Galactic center region are presented in Sec.~\ref{sec:Discussion}. A 
study of the surrounding regions and the dark matter upper limit will be 
discussed in a second publication.

%% ############################################################
%% ############################################################
%% ############ LZA observations
%% ############################################################
%% ############################################################
\section{Large zenith angle observations}
\label{sec:LZA}

The stereoscopic method of shower reconstruction in ground-based 
atmospheric Cherenkov telescopes (such as VERITAS) is based on the 
intersection of the major axes of the parameterized Cherenkov images 
(Hillas) recorded in individual telescopes \citep{Hofmann1999}. In 
general, this method is very powerful since it makes use of the full 
capabilities of the stereoscopic recording of air showers. In the 
following this method is referred to as the {\it geometrical} method.

An alternative technique has been developed for data taken with 
single-telescopes (e.g. Whipple $10 \, \rm{m}$), using an estimate of 
the {\it displacement} parameter which is measured along the major axis 
of the image between the center of gravity of the Hillas ellipse and the 
shower position in the camera system \citep{BuckelyDisp, Whipple_GC, 
MAGIC_DISP}. For $\gamma$-ray showers the {\it displacement} parameter 
has a certain characteristic expectation value (derived from Monte Carlo 
simulations). Its value can be parameterized as a function of the Hillas 
parameters \citep{HillasParameters} of the corresponding image: the 
length $l$, the width $w$, and the amplitude $s$. Throughout this paper 
this method is referred to as the {\it displacement} method.

%-------------
\begin{figure}[t!]

\plotone{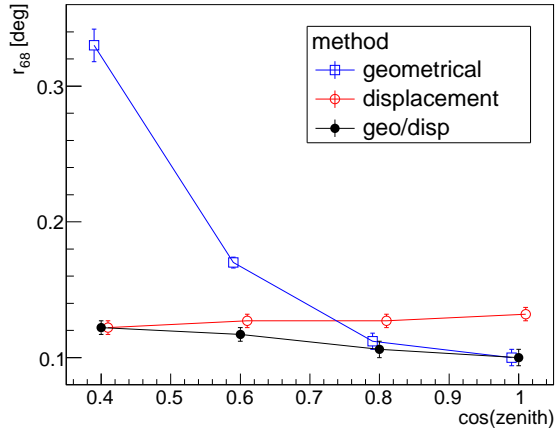}

\caption{\label{fig:LZA_Ang} VERITAS angular resolution ($68\%$ 
containment radius $r_{68}$) as a function of $\cos(\rm{zenith})$ 
derived from Monte Carlo simulations with the requirement of at least 
three images involved in the shower reconstruction. The {\it 
geometrical} algorithm performs well for zenith angles $<40 \deg$ 
($\cos(z) > 0.8$) but gets worse for large zenith angles. At zenith 
angles of $65 \, \deg$, the {\it displacement} method outperforms the 
{\it geometrical} algorithm by a factor of more than $2$. A weighted 
combination of both algorithms ({\it geometrical}/{\it displacement}, 
see text) gives an almost flat angular resolution.}

\end{figure}

Large zenith angle observations are observations where the Cherenkov 
telescopes are pointed to low elevation angles, increasing the average 
distance to the detected showers. This results in a larger footprint for 
the Cherenkov lightpool (increasing the effective area), but a decrease 
of the Cherenkov light intensity also results in an increase of the 
energy threshold. The larger distance to the shower also decreases the 
parallactic displacement between images in the different telescopes. 
Moreover, given the large inclination angle, the angular separation of 
the telescopes projected into the shower plane are foreshortened in one 
dimension. The net effect is a strong reduction in the average stereo 
angle between the major axes of pairs of images, causing a large 
uncertainty in the determination of the {\it geometrical} intersection 
point. This, in turn, leads to a considerable reduction of the angular 
resolution in the reconstruction of the shower direction and impact 
parameter. The {\it displacement} method, on the other hand, does not 
rely on the intersection of axes, making it independent of the stereo 
angle between images. Therefore, no substantial drop in performance is 
expected with increasing zenith angle.

The {\it displacement} parameter as implemented in the VERITAS analysis 
\citep{VER_V407} was parameterized as a function of $l$, $w$, $s$, the 
zenith angle $z$, the azimuth angle $Az$, as well as the pedestal 
variance (a measure for readout noise fluctuations) of the image. In 
contrast to earlier realizations of the method the parameterization is 
done in an orthogonal six-dimensional parameter space stored in the form 
of a lookup table that was trained with an extensive set of Monte-Carlo 
simulations of $\gamma$-ray showers. For each image the {\it 
displacement} parameter is read from the lookup table and results in two 
most likely points of the shower direction with respect to the image 
center of gravity (CoG, in camera coordinates): $\rm{CoG} \pm \rm{\it 
displacement}$ along the major axis of the parameterized image. The 
combination of the points of all images involved in the event resolves 
the two-fold ambiguity\footnote{Therefore, the method requires $N \geq 
2$ images to work.}. The reconstruction of the shower impact parameter 
proceeds in a similar way, again making use of a multi-dimensional 
lookup table.

%-------------
\begin{figure}[t!]

\plotone{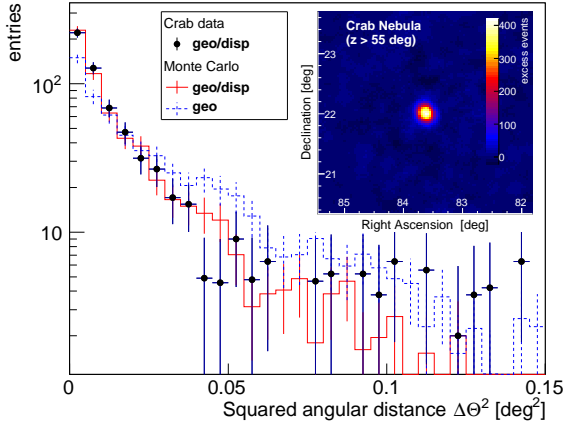}

\caption{\label{fig:LZA_Crab} The data points show the angular 
distribution of excess events from $6.5 \, \rm{hrs}$ of Crab Nebula 
observations taken at zenith angles $z>55 \, \deg$. The showers were 
reconstructed with the combined {\it geometrical}/{\it displacement} 
method. The solid red line represents the angular distribution of Monte 
Carlo events (same reconstruction method) covering the same zenith angle 
range as the data. The dashed blue line shows the distribution of Monte 
Carlo events which were reconstructed with the standard {\it 
geometrical} algorithm. The inlay shows the smoothed excess sky map of 
the Crab Nebula data ({\it geometrical}/{\it displacement} method).}

\end{figure}

Figure~\ref{fig:LZA_Ang} shows the angular resolution of both methods 
({\it geometrical} and {\it displacement}) as a function of the cosine 
of the zenith angle $z$, derived from Monte Carlo simulations.  While 
the angular resolution of the {\it displacement} method remains almost 
independent of $\cos (z)$, the angular resolution of the standard {\it 
geometrical} method becomes increasingly worse at large zenith angles. A 
further improvement is achieved if both methods are combined: $d = 
d_{\rm{geo}} \cdot (1-w') + d_{\rm{disp}} \cdot w'$, with the weight 
being calculated as $w' = \exp(-12.5 \cdot (\cos(z)-0.4)^{2})$ and $w' = 
1$ for $\cos(z) < 0.4$, respectively. Both methods benefit in similar 
ways from an additional requirement of $N \geq 3$ images in the event 
reconstruction. The method was applied to $6.5 \, \rm{h}$ of LZA Crab 
Nebula data (Fig.~\ref{fig:LZA_Crab}). The data are in good agreement 
with the simulations and illustrate the clear improvement the {\it 
displacement} method provides in the case of LZA observations. The 
spectrum reconstructed from the Crab Nebula observations is shown in 
Fig.~\ref{fig:SED} (in Sec.~\ref{sec:GC_VERITAS}) and is found to be in 
reasonable agreement with the H.E.S.S. measurements obtained from lower 
zenith angles. The LZA Crab data set indicates an improvement in 
sensitivity of $30 - 40 \, \%$ when using the combined {\it 
displacement}/{\it geometrical} method compared to the {\it geometrical} 
method alone.

%% ############################################################
%% ############################################################
%% ############ VERITAS observations of the Galactic center
%% ############################################################
%% ############################################################
\section{The Galactic center region imaged by VERITAS} \label{sec:GC_VERITAS}

%%%%%%%%%%%%%%%%%%%%%%%%%%%%%%%%%%%%%%%%%
\begin{figure}[t!]

\plotone{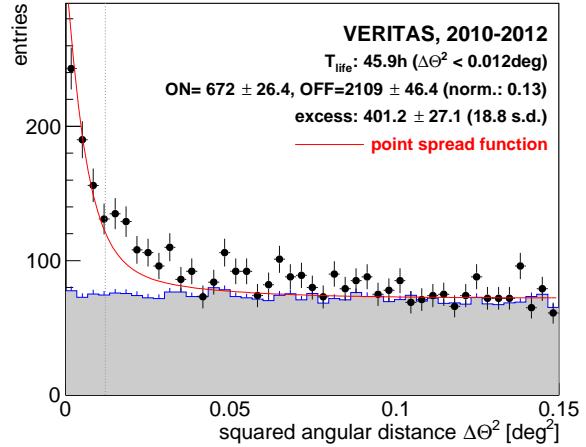}

\caption{Distribution of the squared angular distance $\Delta 
\theta^{2}$ between the reconstructed shower direction and the nominal 
position of the Galactic center (data points). The distribution is also 
shown with respect to the events from the reflected OFF regions (shaded 
area) that are used to determine the background. The red curve 
represents the point-spread function determined from Monte-Carlo 
simulations for the corresponding zenith angle interval, normalized to 
the measured excess determined from the $\Delta \theta^{2} \leq 0.012 \, 
\rm{deg}^{2}$ regime (vertical dotted line).}

\label{fig:ThetaSq}
\end{figure}

VERITAS is an array of four $12 \, \rm{m}$ diameter imaging atmospheric 
Cherenkov telescopes and is located at the base camp of the Fred 
Lawrence Whipple Observatory in southern Arizona at an altitude of $1280 
\, \rm{m}$ \citep{VERITAS}. VERITAS is sensitive to $\gamma$-rays in the 
energy range of $100 \, \rm{GeV}$ to several tens of TeV. For 
observations close to zenith, sources of $10 \%$ ($1 \%$) of the 
strength and spectrum of the Crab Nebula are detected at the level of 
$5$~s.\,d.~in $0.5 \, \rm{hrs}$ ($26 \, \rm{hrs}$), respectively.

%%%%%%%%%%%%%%%%%%%%%%%%%%%%%%%%%%%%%%%%%
\begin{figure*}[t!]

\centering{
\includegraphics[width=0.98\textwidth]{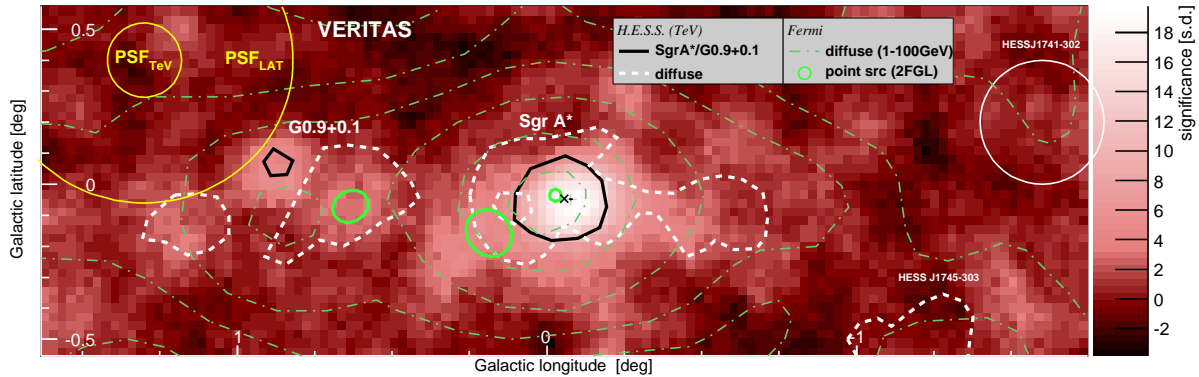}}

\caption{VERITAS sky map of the Galactic center region (smoothed excess 
significances, ring background model). The VERITAS centroid position is 
indicated by the cross (statistical and systematic errors) and the 
Sgr\,A* radio position is indicated by the 'x'. The solid black contour 
lines indicate the Galactic center and the supernova remnant G\,0.9+0.1 
as seen by H.E.S.S. \citep{HESS_SgrA}. The white dashed contour lines 
indicate the H.E.S.S. diffuse emission along the Galactic plane and from 
HESS\,J1745-303 \citep{HESS_SgrA_Diffuse}. The position of 
HESS\,J1741-302 is indicated by the labeled circle. The green solid 
ellipses indicate the positions of the MeV/GeV sources and their $95\%$ 
position errors taken from the second {\it Fermi}/LAT catalog 
\citep{Fermi_SecondCatalog}. The green dash-dotted contour lines 
indicate the $1-100 \, \rm{GeV}$ diffuse emission (after subtracting 
point-sources, Galactic, and extragalactic backgrounds) as measured by 
{\it Fermi}/LAT \citep{Fermi_Contours}. The instruments' point spread 
functions ($68\%$ containment radius) of VERITAS and {\it Fermi}/LAT 
($1-100 \, \rm{GeV}$, calculated for the 2FGL log parabola spectrum of 
the source coincident with SgrA*) are indicated in the upper left 
corner.}

\label{fig:Skymap}
\end{figure*}

The Galactic center was observed by VERITAS in 2010--2012 for $46 \, 
\rm{hrs}$ (good quality data, dead-time corrected). Given the 
declination of the Galactic center, the observations were performed at 
large zenith angles in the range of $z = 60.2-66.4 \deg$, resulting in 
an average energy threshold (energy corresponding to the peak detection 
rate for a Crab-like spectrum) of $E_{\rm{thr}} \simeq 2.5 \, \rm{TeV}$. 
The shower direction and impact parameter were reconstructed with the 
{\it geometrical}/{\it displacement} method as described in 
Sec.~\ref{sec:LZA}. Other than using the {\it displacement} method, the 
standard analysis procedure was applied with event selection cuts {\it 
a-priori} optimized for weak, hard-spectrum sources: angular separation 
between source position and reconstructed shower direction of $\Delta 
\theta^{2} \leq 0.012 \deg^{2}$, mean scaled width/length $\leq 
1.04/1.25$, and $N \geq 3$ images per event.

In the case of ground-based Cherenkov astronomy, the atmosphere acts as 
a calorimeter and its influence on the transmission of Cherenkov light 
is the single largest contributor to the systematic uncertainty in the 
estimate of the reconstructed TeV $\gamma$-ray energy $E$. The overall 
uncertainty for close-to-zenith observations is estimated to be on the 
order of $\Delta E / E \simeq 0.2$, with an atmospheric contribution of 
about $0.15$. For a spectrum $\rm{d}N/\rm{d}E \propto E^{-2.5}$ this 
translates into an error in flux of $\Delta \Phi / \Phi \simeq 0.2$. The 
column density of the atmosphere changes with $1 / \cos(z)$ and, 
conservatively, the systematic error in the energy/flux reconstruction 
can be expected to scale accordingly. For the Galactic center 
observations the contribution of the systematic effect induced by the 
atmosphere therefore roughly doubles as compared to low-zenith angle 
observations and we estimate the systematic error on the LZA flux 
normalization to be $\Delta \Phi / \Phi \simeq 0.4$.

This section describes the results of the TeV $\gamma$-ray source 
coincident with the Galactic center. In the three-year data set, an 
excess on the order of $18 \, \rm{s.\,d.}$~is found at the position of 
Sgr\,A* (see Fig.~\ref{fig:ThetaSq}). The background for the excess 
study (as well as for the energy spectrum, see below) was estimated from 
seven regions placed at the same radial camera distance as the source 
region (reflected background model, see \citet{BG_Models}). The tail in 
the angular excess distribution can likely be explained by a 
contribution from the surrounding diffuse emission 
\citep{HESS_SgrA_Diffuse} which becomes increasingly important at higher 
energies \citep{HESS_UpdatedSpectrum}. More detailed studies on the 
diffuse emission morphology will be presented in a second paper.

Figure~\ref{fig:Skymap} shows the VERITAS sky map of the Galactic center 
region. The background in this figure was estimated using a ring-like 
region ($0.45\deg \leq r \leq 0.7\deg$) surrounding each test position 
(ring background model, see \citet{BG_Models}), with a correction term 
taking into account the camera acceptance. Both background models 
exclude known sources from the background estimate (HESS\,J1741-302, 
HESS\,J1745-303, G\,0.9+0.1, and the Galactic center itself). A fit of 
the point spread function to the uncorrelated excess sky map results in 
a centroid position of the excess in Galactic coordinates of $\rm{long} 
= (-0.077 \pm 0.006_{\rm{stat}} \pm 0.013_{\rm{sys}}) \deg$ and 
$\rm{lat} = (-0.049 \pm 0.003_{\rm{stat}} \pm 0.013_{\rm{sys}}) \deg$ 
with a fit quality of $\chi^{2}/\rm{d.o.f.} = 77.1/61$. We name the 
VERITAS source VER\,J1745-290. This position is compatible with the 
Galactic center position ($\rm{long} = -0.056 \deg$ and $\rm{lat} = 
-0.046 \deg$) and the position measured by H.E.S.S. Both positions are 
indicated in Fig.~\ref{fig:Skymap} which also shows the contour lines of 
the diffuse emission measured by H.E.S.S. \citep{HESS_SgrA_Diffuse}. 
Furthermore, Figure~\ref{fig:Skymap} shows the positions of the MeV/GeV 
sources taken from the 2FGL {\it Fermi} catalog 
\citep{Fermi_SecondCatalog}, as well as the contour lines of the $1-100 
\, \rm{GeV}$ diffuse emission after subtraction of point sources, 
extragalactic, and Galactic backgrounds \citep{Fermi_Contours}.

%%%%%%%%%%%%%%%%%%%%%%%%%%%%%%%%%%%%%%%%%
\begin{figure}[t!]

\plotone{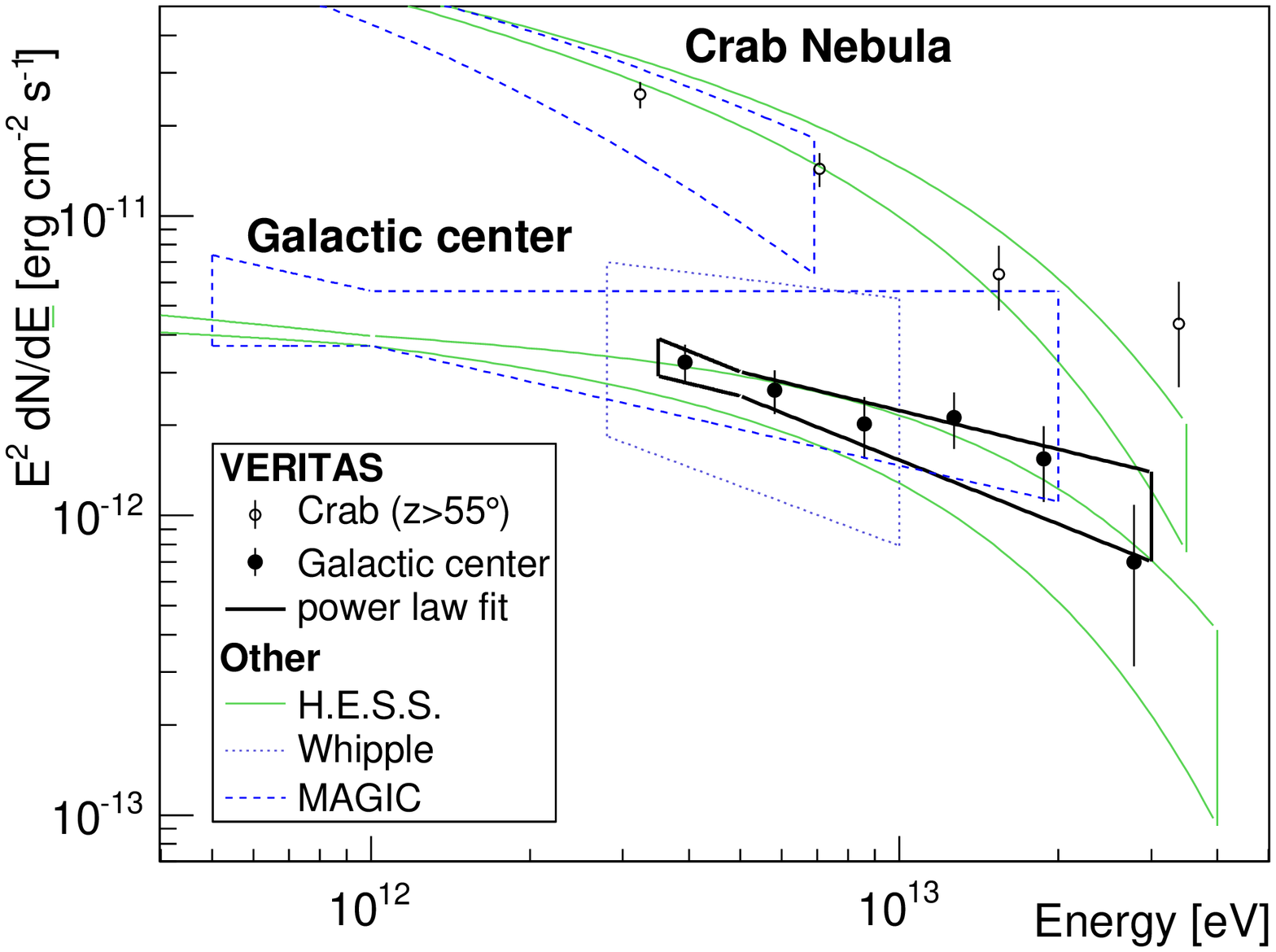}

\caption{VERITAS energy spectra shown for LZA observations of the Crab 
nebula and the Galactic center (statistical errors only). Also indicated 
are the fit results of the spectra measured by Whipple (dotted bow tie, 
\citet{Kosack_Thesis}), H.E.S.S. (solid bow ties, \citet{HESS_Crab, 
HESS_GC_Spectrum}), and MAGIC (dashed bow ties, \citet{MAGIC_Crab, 
MAGIC_GC}).}

\label{fig:SED}
\end{figure}

The VERITAS energy spectrum obtained from the three-year data set is 
shown in Fig.~\ref{fig:SED} and is found to be compatible with the 
spectra measured by Whipple \citep{Kosack_Thesis}, H.E.S.S., and MAGIC. 
It can be described ($\chi^{2}/\rm{dof} = 2.1/4$) by a power law 
$\rm{d}N/\rm{d}E = I_{0} (E/5 \, \rm{TeV})^{-\Gamma}$ with a 
normalization at the decorrelation energy of $5 \, \rm{TeV}$ of $I_{0} = 
(6.89 \pm 0.64_{\rm{stat}} \pm 2.75_{\rm{syst}}) \, 10^{-14} \, \rm{ph.} 
\, \rm{cm}^{-2} \, \rm{s}^{-1} \, \rm{TeV}^{-1}$ and a photon index of 
$\Gamma = 2.57 \pm 0.14_{\rm{stat}} \pm 0.2_{\rm{syst}}$. Since the LZA 
effective area of the VERITAS observations compensates for the shorter 
exposure ($46 \, \rm{h}$) as compared to the low-zenith H.E.S.S. 
observations ($93 \, \rm{h}$), the statistical errors of the VERITAS $E
> 2.5 \, \rm{TeV}$ data points are comparable to those of the 
H.E.S.S.~measurements. Recently, the H.E.S.S.~collaboration reported an 
updated energy spectrum that was corrected for the energy-dependent 
contribution from the surrounding diffuse emission, leading to a lower 
cut-off energy around $10 \, \rm{TeV}$ \citep{HESS_UpdatedSpectrum}.

%%%%%%%%%%%%%%%%%%%%%%%%%%%%%%%%%%%%%%%%%
\begin{table}[t!]

\begin{tabular}{rrr}

year & begin [MJD] & end [MJD] \\
\hline \hline
\noalign{\smallskip}

2010 & 55300.4 & 55308.4 \\
 & 55328.3 & 55334.4 \\
 & 55352.3 & 55366.3 \\
2011 & 55681.4 & 55694.4 \\
 & 55707.3 & 55710.4 \\
 & 55734.2 & 55743.3 \\
2012 & 56033.4 & 56049.5 \\
 & 56063.3 & 56067.4 \\

\end{tabular}

\caption{The time spans (MJD) of the individual observation periods 
(gray data points in Fig.~\ref{fig:LC}). Note, that observations were 
not performed continuously within each period, but were spread out in 
$20-60 \, \rm{min}$ data segments in individual nights.}

\label{tab:SeasonTimes}

\end{table}

\begin{figure}[t!]

\plotone{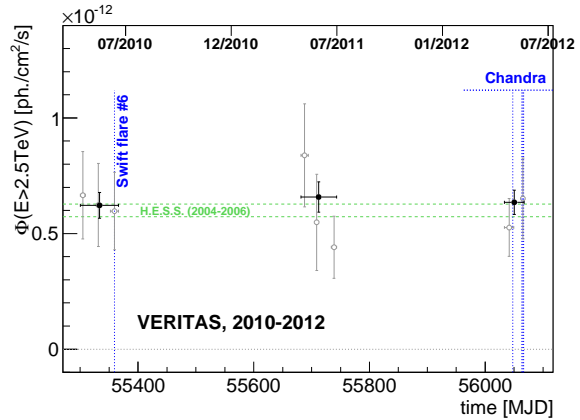}

\caption{Integral flux above $2.5 \, \rm{TeV}$ from the direction of the 
Galactic center on a month-by-month basis (open points), as well as the 
yearly averages (solid points). The time of the X-ray flare \#6 detected 
by Swift \citep{Swift_X-rayFlares} is indicated. The period of X-ray 
Chandra observations of Sgr\,A* \citep{Chandra2012} is shown as the 
dotted horizontal line. VERITAS synchronized four observations with the 
Chandra pointings (same day, vertical lines). The dashed horizontal 
lines indicate the statistical error range of the 2004-2006 average TeV 
$\gamma$-ray flux which was derived by integrating the H.E.S.S. spectrum 
shown in Fig.~\ref{fig:SED} for $E > 2.5 \, \rm{TeV}$.}

\label{fig:LC}
\end{figure}

The night-by-night integral fluxes above $2.5 \, \rm{TeV}$ are 
calculated by folding a fixed spectral slope (derived from the energy 
spectrum for the full data set: $\rm{d}N/\rm{d}E \propto E^{-2.6}$) with 
the effective area for the zenith angle of the corresponding night and 
comparing it with the excess counts above the threshold found in the 
data. The fluxes were binned according to observation periods of $\sim$3 
week duration (as summarized in Tab.~\ref{tab:SeasonTimes}) and are 
shown in Fig.~\ref{fig:LC} together with the yearly integral fluxes 
obtained from integrating the reconstructed energy spectra obtained for 
the individual years. No evidence for flux variability was found in the 
three-year data (the fit of a constant function to the run-by-run light 
curve (20 \, \rm{min} segments) yields a fit quality of 
$\chi^{2}/\rm{dof} = 117/150$). The H.E.S.S.~collaboration reported a 
fit quality of $\chi^{2}/\rm{dof} = 233 / 216$ as a result of a 
comparable study based on their 2004-2006 data set divided into $28 \, 
\rm{min}$ segments \citep{HESS_GC_Spectrum}.

%%%%%%%%%%%%%%%%%%%%%%%%%%%%%%%%%%%%%%%%%

%% ############################################################
%% ############################################################
%% ############ Discussion
%% ############################################################
%% ############################################################
\section{Discussion} \label{sec:Discussion}

%% ############################################################
\subsection{VERITAS results in the context of multi-wavelength data}

The following potential counterparts are 
located\footnote{http://simbad.u-strasbg.fr} within the $1 \, \arcmin$ 
position uncertainty (statistical plus systematic) of the VERITAS 
excess: (i) The Galactic center Sgr\,A*, (ii) the supernova remnant 
Sgr\,A~East, (iii) the PWN G\,359.95-0.04, (iv) the low mass X-ray 
Binary AX\,J1745.6-2901, (v) nine maser objects, and (vi) about $150$ 
X-ray sources. Here, (i)-(iii) have the highest potential for TeV 
$\gamma$-ray emission. Sgr\,A~East, however, was excluded as TeV 
counterpart by H.E.S.S. \citep{HESS_GC_Location}. PWN G\,359.95-0.04 is 
discussed as potential counterpart in Sec.~\ref{subsec:ModelComparison}. 
In general, a contribution to the measured TeV $\gamma$-ray flux from 
more than one object cannot be excluded.

Sgr\,A* is known to exhibit $2-10 \, \rm{keV}$ X-ray flares above the 
quiescent level on a regular basis, as for example observed in the 
2006--2011 Swift monitoring campaign \citep{Swift_X-rayFlares}. A bright 
X-ray flare (flare \#6, MJD 55359) was detected by Swift during the 2010 
VERITAS monitoring but no TeV data were taken on that particular night. 
The time of the X-ray flare is indicated in Fig.~\ref{fig:LC} and its 
spectrum is shown in the spectral energy distribution (SED) in 
Fig.~\ref{fig:SED2} together with a baseline measurement of the 
continuum emission from the extended region surrounding the Galactic 
center (including the contribution of Sgr\,A*). High spatial resolution 
X-ray observations were conducted in 2012 by Chandra \citep{Chandra2012} 
for a total of $3 \, \rm{Ms}$ leading to the detection of 39 X-ray 
flares from the Galactic center with durations ranging from 
$O(100\,\rm{s})$ to $O(8 \, \rm{ks})$. The observed flare luminosities 
in the $2-10 \, \rm{keV}$ band reached $10^{34}$ to $2 \cdot 10^{35} \, 
\rm{ergs}/\rm{s}$. The time span of the Chandra campaign is indicated in 
Fig.~\ref{fig:LC}, where four nights had quasi-simultaneous coverage 
(same night) by VERITAS. Medium-intensity X-ray flares were detected in 
two out of those four nights at MJD\,56047.7 and 56066.4. However, no 
significant flux changes were observed in the TeV band (see 
Sec.~\ref{subsec:VER_VarEstimate} for an estimate of the VERITAS 
sensitivity to detect variability). In an earlier campaign an X-ray 
flare was observed during joint H.E.S.S./Chandra observations in 2005; 
but no increase in TeV $\gamma$-ray flux was measured 
\citep{HESS_ChandraCampaign}.

One possible origin of the observed X-ray flares is a change/disruption 
in accretion rate. In models where the TeV emission comes from particle 
acceleration near the black hole, one might expect some connection 
between the variation in the X-ray emission of the accretion and the TeV 
$\gamma$-ray production (see next section). The frequency of X-ray 
flares exceeding the quiescent state by a factor of 10 is estimated to 
be roughly one flare per day \citep{Chandra2012}. The frequency of 
bright ($L_{\rm{X}} > 10^{35} \, \rm{ergs}/\rm{s}$) X-ray flares is 
estimated to be $0.1-0.2$ per day \citep{Swift_X-rayFlares}. Given (i) 
the sensitivity of the VERITAS LZA Galactic center observations, (ii) 
the X-ray flare intensity and (iii) flare frequency, it is challenging 
to correlate (with either direct or delayed response functions) the two 
wave bands on the basis of individual flares unless much stronger X-ray 
flares are observed ($\gg 10$ times the X-ray quiescent level). Most TeV 
emission models (see next section) predict a 'smoothing out' of the 
accretion/flare activity in the TeV response, if there is any 
relationship at all.

Four more medium-intensity X-ray flares with durations of less than one 
hour and energies reaching up to $79 \, \rm{keV}$ were detected in a 
NuSTAR campaign conducted in summer/fall 2012 \citep{NuStar2012}. 
Although the VERITAS 2012 observations had already ended by that time, 
the measured spectra of the two strongest flares J21$_{2}$ (07/2012) and 
O17 (10/2012) are shown in Fig.~\ref{fig:SED2} for reference.

At current times the emission level from the Galactic center is roughly 
10 orders of magnitude below its Eddington luminosity 
\citep{Fluorescense1, Fluorescense3}. Spatial and temporal variations in 
the X-ray flux measured from molecular clouds surrounding the Galactic 
center have been found \citep{Fluorescense1, Fluorescense2, 
Fluorescense3}. The results are interpreted as a bright $10^{39} \, 
\rm{ergs}/\rm{s}$ outburst of a source coincident with Sgr\,A* that 
happened $O(100) \, \rm{y}$ ago. These findings indicate long-term 
variations in accretion/brightness of the central source 
\citep{Fluorescense2}. Recently, the Fermi/LAT instrument discovered two 
large $\gamma$-ray bubbles extending below and above the Galactic center 
\citep{FermiBubbles}. Although the origin of the bubbles remains unclear 
so far, a significant increase in energy injection from the Galactic 
center on times scales of Myr is discussed as one of the possibilities 
\citep{FermiBubbles}; for example, in the form of a plasma jet 
originating from the (previously brighter) active galactic nucleus in 
our galaxy \citep{FermiBubbles_AGN, FermiBubbles_AGN2}.

%% ############################################################
\subsection{Comparison to models} \label{subsec:ModelComparison}

\begin{figure*}[t]

\centering{
\includegraphics[width=0.98\textwidth]{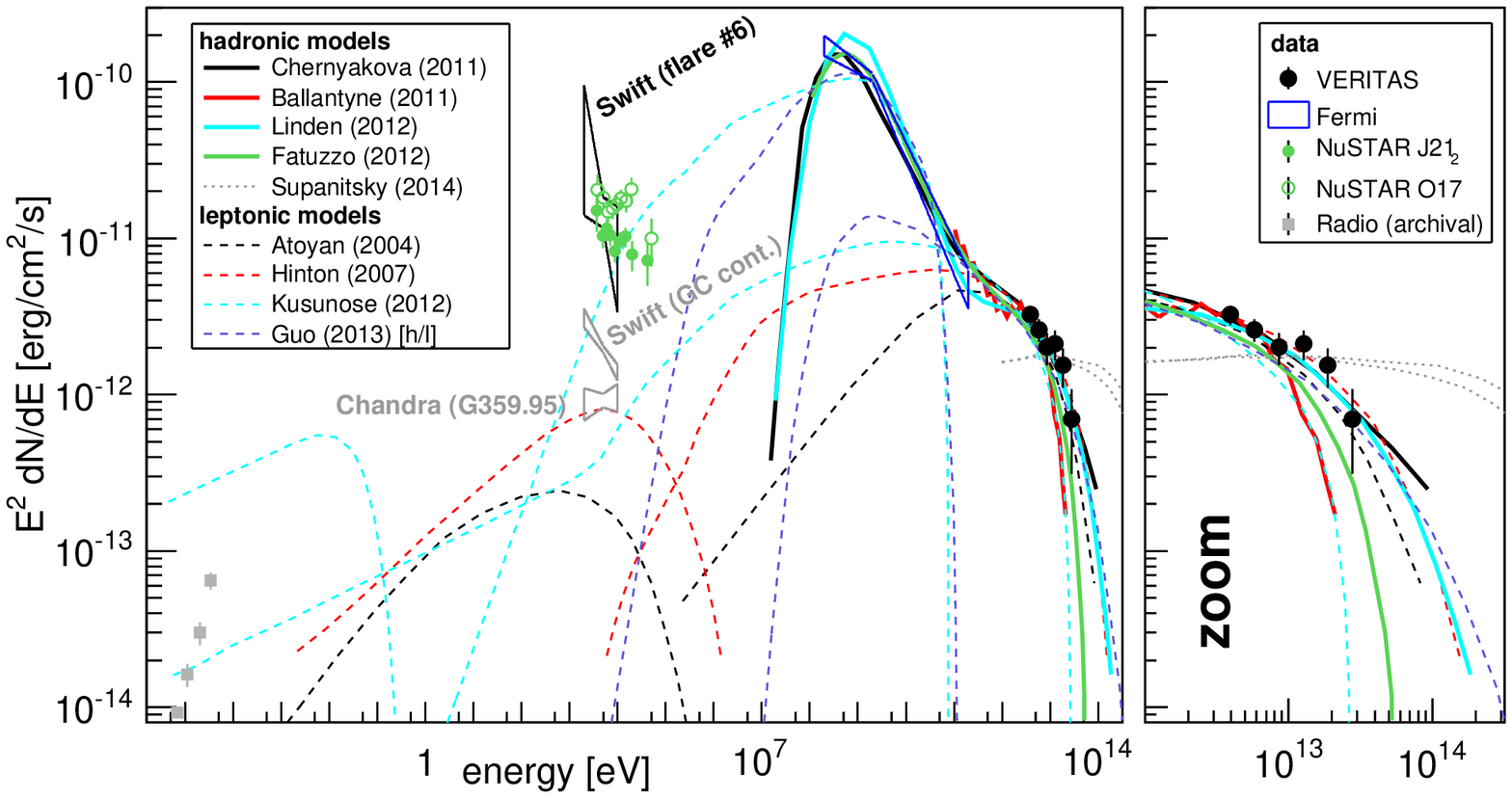}}

\caption{{\bf Left:} VERITAS spectral energy distribution of the 
Galactic center point-source compared to hadronic and leptonic emission 
models as discussed in the text. The {\it Fermi}/LAT bow tie is taken 
from \citet{Chernyakova2011}. The multi-wavelength data points shown are 
archival radio \citep{SgrA_Radio_SED}, X-ray flares observed by Swift 
\citep{Swift_X-rayFlares} and NuSTAR \citep{NuStar2012}, as well as flux 
measurements from Chandra \citep{GC_Plerion}. {\bf Right:} Zoomed 
version of the TeV regime.}

\label{fig:SED2}
\end{figure*}

A variety of astrophysical models have been proposed to explain the 
GeV/TeV $\gamma$-ray emission from the Galactic center region. This 
section discusses a selection of leading models that cover the range of 
viable hypotheses. The models are shown together with the VERITAS and 
multi-wavelength data in Fig.~\ref{fig:SED2}. Most of the models were 
tuned based on the H.E.S.S.~results so that a general agreement with the 
VERITAS spectrum is not surprising. While some of these models link 
accretion onto the black hole to the X-ray and $\gamma$-ray data, most 
of them find a way to address the lack of variability in the TeV 
emission, and a direct flux correlation between the X-ray/TeV band is 
not predicted. With respect to the TeV $\gamma$-ray emission, models can 
be divided into two broad classes: hadronic models or leptonic models 
depending on which species of energetic particles dominates the 
$\gamma$-ray emission.

%%%%%%%%%%%%%%%%%%%%%%%%%%%%%%%%%%%%%%%%%

%%----------------------------
\subsubsection{Hadronic models} 

Most of the hadronic acceleration models such as \citet{Chernyakova2011} 
and \citet{Ballantyne2011} explain the emission by the following 
mechanism: (i) protons are being accelerated in the black hole vicinity 
at distances of up to a few tens of Schwarzschild radii. (ii) The 
accelerated protons diffuse out into the interstellar medium where they 
(iii) undergo nuclear interactions and produce neutral pions which decay 
into GeV/TeV $\gamma$-rays: $\pi^{0} \rightarrow \gamma \gamma$. The 
spectral break between the MeV/GeV and TeV spectra is explained by a 
strong energy dependence of the diffusion coefficient separating the 
high-energy particles into two different diffusion regimes 
\citep{Chernyakova2011}. Changes in the TeV flux can potentially be 
caused by changes in the black hole vicinity (e.g. accretion) but will 
not manifest themselves instantaneously. The time scales of flux 
variations in these models are $\sim$$10^{4} \, \rm{y}$ at MeV/GeV 
energies (old flares) and $\sim$$10 \, \rm{y}$ at $E > 10 \, \rm{TeV}$ 
(`new' flares caused by recently injected high-energy particles). 
Significant spectral variability has not been seen but is also not 
strongly constrained following $\sim$15 years of observations by 
Whipple, H.E.S.S., MAGIC, and VERITAS. However, such variability can be 
expected in this model for $E > 10 \, \rm{TeV}$ with the TeV spectrum 
softening following an outburst \citep{Ballantyne2011}. Constraining the 
$E > 10 \, \rm{TeV}$ spectral variability would serve as an important 
test for this class of models (see Sec.~\ref{subsec:VER_VarEstimate} for 
an estimate of the VERITAS sensitivity to constrain flux variability). 
\citet{Linden2012} discuss the surrounding gas as a proton target that 
defines the morphology of the TeV $\gamma$-ray emission.

\citet{DiffuseCR_Acceleration} interpret the inflected structure in the 
GeV/TeV spectrum as a hint for an energizing process more complicated 
than typical non-relativistic diffusive shock acceleration. While 
questioning some of the assumptions made by \citet{Chernyakova2011} 
their model assumes a steady-state cosmic-ray ejection by Sgr\,A* 
without a particle diffusion coefficient that strongly depends on 
energy. Their model treats the inner parsecs of the galaxy as a uniform 
{\it wind zone} (interactions of stellar winds from the surrounding 
young stars) that encompasses a high-density {\it molecular torus} with 
an inner radius of $1.2 \, \rm{pc}$ and a thickness of $1 \, \rm{pc}$. 
The high-energy tail of the thermal proton distribution near the black 
hole serves as a seed population for the stochastic acceleration 
process. The $\gamma$-rays are produced via $\pi^{0}$ decays or 
electromagnetic $\pi^{\pm}$ cascades as a consequence of pp scattering. 
In this scenario the emission observed in the {\it Fermi}/LAT band is 
dominated by scattering in the torus whereas the TeV emission is 
dominated by scattering in the wind zone.

Motivated by the recent IceCube detection of five $E>30 \, \rm{TeV}$ 
neutrinos from the direction of the Galactic center region 
\citep{IceCubeNeutrinos}, \citet{Supanitsky2014} studies the interaction 
of cosmic ray protons (i) accelerated by sources in the Galacic center 
region, (ii) interacting with ambient protons, and (iii) calculates the 
resulting $\gamma$-ray and neutrino spectra. It should be stressed that 
\citet{Supanitsky2014} discusses a hypothetical PeV cosmic ray 
accelerator (Pevatron) located in the Galactic center region, which has 
no experimental evidence for its existence so far. Not surprisingly, the 
predicted spectrum differs substantially from the models discussed above 
and can be constrained by more sensitive observations at the highest 
energies.

%%----------------------------
\subsubsection{Leptonic models} 

\citet{Atoyan2004} discuss a black hole plerion model. Here, a 
magnetized leptonic wind originates from the advection dominated 
accretion flow surrounding the black hole and results in a termination 
shock located at a distance of $3 \times 10^{16} \, \rm{cm}$ ($\simeq 
7500$ Schwarzschild radii) from the black hole. The shock accelerates 
leptons to relativistic energies which in turn produce TeV $\gamma$-rays 
via inverse Compton scattering. This model fails to explain the flux in 
the MeV/GeV regime (Fig.~\ref{fig:SED2}). However, given the limited 
angular resolution, the emission observed in the {\it Fermi}/LAT band 
may well originate from a different region, different source, or a 
different spectral component in the same source. Future {\it 
Fermi}/VERITAS flux correlation studies will serve as crucial 
experimental inputs to understand a possible common versus separate 
origin of these two SED components. The hadronic models, on the other 
hand, can explain the MeV/GeV part of the SED by the superposition of 
different flare stages that occured in the recent history of the source. 
The flux variability time scale in \citet{Atoyan2004} is on the order of 
$T_{\rm{var}} \sim$$100 \, \rm{y}$ and therefore provides a prediction 
that would be falsified by the detection of TeV $\gamma$-ray flux 
variability.

%R_schw = 13.3 Mio km = 13.3 e6 km = 13.3 e6 e5 cm = 1.33 e12 cm

\citet{Kusunose2012} discuss a leptonic model that involves a different 
location/mechanism for the MeV/GeV versus TeV emission. The observed 
(quasi-continuous) X-ray flaring is seen as synchrotron emission of 
non-thermal electrons that are injected and accumulate in a region of $r 
\leq 10^{18} \, \rm{cm}$ ($\simeq 7.5 \cdot 10^{5}$ Schwarzschild radii) 
around the black hole. These electrons produce MeV/GeV $\gamma$-rays 
seen by {\it Fermi}/LAT via inverse Compton scattering off soft 
star/dust photons\footnote{The model versus data difference in the radio 
regime is explained by a difference in emission regions considered 
versus measured.}. By increasing the electron Lorentz factor and 
reducing the injection rate the model can be tuned to describe the TeV 
emission, as well. However, it cannot describe the MeV/GeV and TeV 
emission with a single set of model parameters suggesting different 
origins or emission zones of the two measured spectral components.

\citet{Hinton2007} link the TeV emission to the recently discovered 
pulsar wind nebula G\,359.95-0.04 \citep{GC_Plerion} located only $0.3 
\, \rm{pc}$ (projected) away from the Galactic center. The authors 
adjust the PWN/TeV scenario to the very high density of low-frequency 
radiation found in the particular region of the Galactic center. This 
environment leads to a hardening of the high-energy electron spectrum 
and to more efficient TeV emission as compared to the same PWN located 
in a `regular' environment. The model does not describe the {\it 
Fermi}-observed MeV/GeV spectrum which in this scenario would originate 
from a different location. Given the instruments' point spread functions 
(VERITAS: $\simeq 0.1 \deg$, {\it Fermi}/LAT: $\simeq 0.5 \deg$ at 
$1-100 \, \rm{GeV}$, $\simeq 2.5 \deg$ at $100 \, \rm{MeV} - 1 \, 
\rm{GeV}$), neither {\it Fermi} nor VERITAS is capable of distinguishing 
between the positions of G\,359.95-0.04 and the Galactic center based on 
the measured excess location.

It should be noted that the $E > 10 \, \rm{TeV}$ leptonic emission in 
the models discussed above is strongly Klein-Nishina suppressed in the 
case of photon fields with temperatures above $100 \, \rm{K}$.

%%----------------------------
\subsubsection{Hybrid models} 

\citet{TeV_Hybrid2013} discuss a hybrid model that assumes simultaneous 
acceleration of hadrons and electrons during a past phase of activity of 
the Galactic center and significant contributions to the observed SED by 
both hadronic and leptonic radiative processes. The particles are 
accelerated in a region surrounding the black hole with a radius of 
approximately 10 Schwarzschild radii, diffuse outward and interact with 
interstellar gas and radiation fields, respectively. In this scenario 
the hadrons are responsible for the TeV emission via the $\pi^{0}$ decay 
channel from a region with $r < 3 \, \rm{pc}$. Fast cooling electrons, 
on the other hand, would dominate the MeV/GeV emission via inverse 
Compton scattering off the soft background photons in a region with $r < 
1.2 \, \rm{pc}$. The cut-off in the MeV/GeV spectrum moves towards lower 
energies and the spectrum softens with increasing time since particle 
injection activity. The time dependence of the TeV spectrum is weaker, 
with a softening trend in time. Both spectral components drop in flux by 
roughly a factor of two within $O(100 \, \rm{y})$. Within this 
framework, the measured MeV/GeV and TeV spectra can be simultaneously 
explained by a $10^{48} \, \rm{erg}$ injection event roughly 200 years 
ago. The authors note that an outburst similar to the one observed by 
Chandra in 2012 would lead to TeV emission three orders of magnitude 
below the current measurements~-- implying that much stronger past 
activity was responsible for the current state of MeV/GeV/TeV emission.

%% ############################################################ 
\subsection{Prospects for dark matter limits}

A number of extensions to the standard model of particle physics predict 
new particles with TeV-scale masses.  Supersymmety, e.g., provides a 
natural candidate for WIMP dark matter, the neutralino or lightest 
(stable) sypersymmetric particle. If these WIMPs were thermal relics, 
their interactions in the early Universe imply that they will interact 
with ordinary matter in the present, annihilating to form standard model 
particles and $\gamma$-rays or, in some cases, even decaying.

In almost any scenario of cold dark matter structure formation, the 
Milky Way halo is thought to be peaked in the Galactic center region and 
the annihilation rate, proportional to the density squared, would be 
even more strongly peaked near the Galactic center. WIMPS could 
annihilate directly to $\gamma$-rays forming narrow lines (through $\chi 
\chi \rightarrow \gamma \gamma$ or $\chi \chi \rightarrow \gamma + 
Z^{0}$) or annihilate to quarks or heavy leptons, hadronizing and 
producing secondary $\gamma$-rays in a continuum \citep{Neutralino}. The 
resulting spectrum would have a cut-off near the WIMP mass $m_{\chi}$, 
as well as a detailed spectral shape determined by the annihilation 
channel.

Prior to the LHC, the natural mass for WIMPs was thought to fall below 
TeV energies, the energy range previous searches had been focusing on 
(see e.g. \citet{HESS_GC_DM}). But with recent contraints from the LHC, 
multi-TeV scale WIMPs have attracted increasing attention 
\citep{HighMass_DM}. Above a few TeV, nonperturbative effects (e.g., 
Sommerfeld enhancements from W or Z exchange) could boost the 
annihilation cross-section by more than an order of magnitude.  Thus, 
multi-TeV measurements of $\gamma$-ray emission from the Galactic center 
are of great interest. The study of diffuse emission and upper limits on 
a dark matter annihilation signal will follow this paper that first 
identifies astrophysical point sources.

%% ############################################################ 
\subsection{Prospects of future VERITAS observations} \label{subsec:VER_VarEstimate}

As discussed in Sec.~\ref{subsec:ModelComparison}, most of the emission 
models start to differ in the cut-off regime around $10 \, \rm{TeV}$ 
(see Fig.~\ref{fig:SED2}). Furthermore, some of the hadronic models 
predict variability in the $E>10 \, \rm{TeV}$ flux on time scales of 
$O(10 \, \rm{y})$ whereas the leptonic model family predicts flux 
changes on time scales not shorter than $O(100 \, \rm{y})$. The 
differences are to a large extent the result of different assumptions 
concerning the acceleration rates (and sizes of the emission regions): 
the hadronic models assume abrupt changes in acceleration whereas the 
leptonic models assume much slower variations in the acceleration rate. 
Future VERITAS observations would help to constrain the different models 
by having a more accurate measurement of the cut-off energy, as well as 
better constraints on the time variability of the emission.

In the data set presented in this paper, VERITAS detects emission from 
the direction of the Galactic center above $10 \, \rm{TeV}$ at a 
significance level of $7.5 \, \rm{s.\,d.}$~with a rate of $1.1 \, 
\rm{s.\,d.}$~per $\sqrt{\rm{h}}$. Assuming a continuation of the 
monitoring of $15 \, \rm{h}$ per year the change in $E>10 \, \rm{TeV}$ 
flux can be constrained as follows. Assuming an increase in the flux of 
$0/50/100 \%$ the VERITAS detection within individual years would result 
in excess significances of $4.4/6.7/8.8 \, \rm{s.\,d.}$, respectively. 
The doubling of the flux from one year to another could be detected at 
the level of $3.4 \, \rm{s.\,d}$. An increase by a factor of three would 
be highly significant ($5 \, \rm{s.\,d.}$). An estimate of the 
corresponding sensitivity for flux changes in the whole energy range 
covered by the VERITAS observations ($E \gtrsim 2.5 \, \rm{TeV}$) would 
result in the detection of a $40\%$ increase in flux from one year to 
another at the level of $5.5 \, \rm{s.\,d.}$

The X-ray flares observed by Swift, Chandra, and NuSTAR can reach flux 
amplitudes of more than one order of magnitude above the X-ray 
quiescence level. Although the TeV emission models discussed in 
Sec.~\ref{subsec:ModelComparison} do not predict a direct link, it is 
important to constrain/exclude X-ray/TeV flux correlations (e.g. X-ray 
flares may mark a different mechanism that potentially could be 
accompanied by direct TeV $\gamma-ray$ emission). Assuming a comparable 
increase in X-ray vs.~TeV flux, VERITAS would be able to establish the 
corresponding flux variability at a high level of significance. However, 
the X-ray flares only last for short time scales of $O(1 \, \rm{h})$ so 
that exactly simultaneous X-ray/TeV observations are required during a 
strong X-ray flare to test a possible correlation.

Another strong motivation for a continuation of the TeV monitoring of 
the Galactic center region is the gaseous object G\,2 heading towards 
its center \citep{BH_eats_MC}. Although the predictions for the changes 
in accretion rate vary (but will be further constrained by ongoing 
multi-wavelength campaigns in the years to come), this event marks a 
unique opportunity in which well-defined changes of the environment 
conditions of the black hole vicinity can be used to study the 
corresponding impacts on non-thermal emission in the X-ray band and up 
to MeV/GeV/TeV energies. Although the TeV flux will in most models react 
to changes in accretion on time scales $\gg 1 \, \rm{y}$, short-term 
changes in the high-energy regime by local shock acceleration due to the 
merging process cannot be excluded. Observations in the GeV/TeV regime 
with the next-generation instrument, the Cherenkov Telescope Array 
\citep{CTA}, will be almost one order of magnitude more sensitive and 
will be well-suited for more detailed variability studies.

As of the beginning of 2014 the ongoing radio monitoring with the VLA 
did not yet reveal a significant brightening of the Galactic center 
region due to the merger \citep{VLA_Monitoring}. However, the process of 
merging is believed to last for several decades, with inaccurate 
predictions so far about its exact onset.

%% ############################################################
%% ############################################################
%% ############ Summary and Conclusions
%% ############################################################
%% ############################################################
\section{Summary and conclusions}

The implementation of the {\it displacement} method in the VERITAS data 
analysis chain has substantially improved the shower reconstruction and 
sensitivity for data taken at large zenith angles. It allows detection 
of the Galactic center at the level of $5 \, \rm{s.\,d.}$~in roughly $3 
\, \rm{hrs}$ with $z>60 \, \deg$ observations. The measured energy 
spectrum is found to be in agreement with earlier measurements by 
H.E.S.S., MAGIC, and Whipple. At energies above $2.5 \, \rm{TeV}$ the 
VERITAS measurements are competitive with H.E.S.S. Further constraints 
on emission models can be placed by future observations to measure the 
cut-off energy in the spectrum and to determine limits on the flux 
variability at the highest energies. The recently discovered gaseous 
object G\,2 heading towards the immediate vicinity of the Galactic 
center black hole \citep{BH_eats_MC} represents further motivation for 
future TeV $\gamma$-ray monitoring of this region: in addition to the 
potential for discoveries, the observations will establish a base line 
TeV $\gamma$-ray flux and spectrum that can be used to study possible 
changes caused by the merging process that can potentially last for 
several decades \citep{G2_Simulations, G2_Simulations2, 
G2_Simulations3}. An upper limit on diffuse $\gamma$-ray emission 
surrounding the Galactic center region and, in consequence, a limit on 
the photon flux initiated by the annihilation of dark matter will be 
presented in a separate publication.

%% ############################################################
%% ############ Acknowledgements
%% ############################################################
\acknowledgements

This research is supported by grants from the U.S. Department of Energy 
Office of Science, the U.S. National Science Foundation and the 
Smithsonian Institution, by NSERC in Canada, by Science Foundation 
Ireland (SFI 10/RFP/AST2748) and by STFC in the U.K. We acknowledge the 
excellent work of the technical support staff at the Fred Lawrence 
Whipple Observatory and at the collaborating institutions in the 
construction and operation of the instrument.

%% ############################################################
%% ############ Bibliography
%% ############################################################


\begin{thebibliography}{} % Use for 10-99 references

\bibitem[Abarca et al.~(2014)]{G2_Simulations} Abarca, D., Sadowski, A., 
Sironi, L. 2014, MNRAS, 440, 1125

\bibitem[Abdo et al.~(2010)]{Fermi_FirstCatalog}Abdo, A.A., et al. 2010, 
ApJS, 188, 405

\bibitem[Abramowski et al.~(2011)]{HESS_GC_DM}Abramowski, A., et al. 
2011, Phys. Rev. Lett., 106, 161301

\bibitem[Acero et al.~(2010)]{HESS_GC_Location}Acero, F., et al. 2011, 
MNRAS, 402, 1877-1882.

\bibitem[Aharonian et al.~(2004)]{HESS_SgrA}Aharonian, F.A., et al. 
2004, A\&A, 425, L13

\bibitem[Aharonian et al.~(2006a)]{HESS_SgrA_Diffuse}Aharonian, F.A., et 
al. 2006a, Nature, 439, 695

\bibitem[Aharonian et al.~(2006b)]{HESS_Crab}Aharonian, F.A., et al. 
2006b, A\&A, 457, 899

\bibitem[Aharonian et al.~(2009)]{HESS_GC_Spectrum}Aharonian, F.A., et 
al. 2009, A\&A, 503, 817

\bibitem[Albert et al.~(2006)]{MAGIC_GC}Albert, J., Aliu, E., Anderhub, 
H., et al. 2006, ApJ, 638, L101

\bibitem[Albert et al.~(2008)]{MAGIC_Crab}Albert, J., et al. 2008, ApJ, 
674, 1037

\bibitem[Aliu et al.~(2012)]{VER_V407}Aliu, E., et al. 2012, ApJ, 
754, 77

\bibitem[Aartsen et al.~(2013)]{IceCubeNeutrinos}Aartsen, M., et al.
2013, Science, 342, 1242856

\bibitem[Atoyan \& Dermer (2004)]{Atoyan2004}Atoyan, A., \& Dermer, C.D. 
2004, ApJ, 617, L123

\bibitem[Ballantyne et al.~(2011)]{Ballantyne2011}Ballantyne, D.R., 
Schumann, M., \& Ford, B. 2011, MNRAS, 410, 1521

\bibitem[Ballone et al.~(2014)]{G2_Simulations2} Ballone, A., 
Schartmann, M., Burkert, A., et al. 2014, IAUS, 303, 307

\bibitem[Barriere et al.~(2014)]{NuStar2012}Barriere, N.M., Tomsick, 
J.A., Baganoff, F.K., et al. 2014, (submitted), arXiv:1403.0900

\bibitem[Beilicke et al.~(2011)]{Beilicke2011}Beilicke, M., et al. 
(VERITAS collaboration) 2011, Proc.~2011 Fermi Symp., arXiv 1109.6836

\bibitem[Berge et al.~(2007)]{BG_Models} Berge, D., Funk, S. \& Hinton, 
J. 2007, A\&A, 466, 1219

%\bibitem[Bergstr{\"o}m et al.~(1998)]{GammasFromNWF}Bergstr{\"o}m, L., 
%Ullio, P., \& Buckley, J.H. 1998, APh, 9, 137 

\bibitem[Buckley et al.~(1998)]{BuckelyDisp}Buckley, J.H., Akerlof, 
C.W., Carter-Lewis, D.A., et al. 1998, A\&A, 329, 639

\bibitem[Chandler \& Sjouwerman (2014)]{VLA_Monitoring}Chandler, C.J., 
\& Sjouwerman, L.O. 2014, The Astronomer's Telegram \#5727, see also 
https://science.nrao.edu/science/service-observing

\bibitem[Chernyakova et al.~(2011)]{Chernyakova2011}Chernyakova, M., 
Malyshev, D., Aharonian, F.A., Crocker, R.M., \& Jones, D.I. 2011, ApJ, 
726, 60

\bibitem[CTA Consortium (2010)]{CTA}CTA Consortium 2010, arXiv:1008.3703

\bibitem[Davies et al.~(1976)]{SgrA_Radio_SED}Davies, R.D., Walsh, D.,
\& Booth, R.S. 1976, MNRAS, 177, 319

\bibitem[Degenaar et al.~(2012)]{Degenaar2012} Degenaar, N., Wijnands, 
R., Cackett, E.M., et al. 2012, A\&A, 545, 49

\bibitem[Degenaar et al.~(2013)]{Swift_X-rayFlares}Degenaar, N., Miller, 
J.M., Kennea, J., et al. 2013, ApJ, 769, 155

\bibitem[Domingo-Santamaria et al.~(2005)]{MAGIC_DISP} 
Domingo-Santamaria, E., Flix, J., Rico, J., Scalzotto, V., and Wittek, 
W. 2005, Proc. of the ICRC, 5, 363

\bibitem[Egberts et al.~(2013)]{HESS_UnresolvedDiffuse} Egberts, K., 
Brun, F., Casanova, S., et al. 2013, Proc. of the ICRC 2013, see 
arXiv:1308.0161

\bibitem[Fatuzzo \& Melia (2012)]{DiffuseCR_Acceleration}Fatuzzo, M., 
\& Melia, F. 2012, ApJ, 757, 16

\bibitem[Gillessen, et al.~(2012)]{BH_eats_MC}Gillessen, S., Genzel, R., 
Fritz, T., et al. 2012, Nature 481, 51

\bibitem[Guo et al.~(2012)]{FermiBubbles_AGN2} Guo, F., Mathews, W.G. 
2012, ApJ, 756, 181

\bibitem[Guo et al.~(2013)]{TeV_Hybrid2013}Guo, Y.-Q., Yuan, Q., Liu, 
C., Li, A.-F. 2013, JPhG, 40f5201G

\bibitem[Hartman et al.~(1999)]{Egret_GC}Hartman, R.C., Bertsch, D.L., 
Bloom, S.D., et al. 1999, ApJS, 123, 79

\bibitem[Hillas (1985)]{HillasParameters}Hillas, A.M. 1985, in NASA 
Goddard Space Flight Center 19$^{\rm{th}}$ Int. Cosmic Ray Conf., 
Vol. 3, 445

\bibitem[Hinton et al.~(2008)]{HESS_ChandraCampaign}Hinton, J.A., 
Vivier, M., B\"uhler, R., et al. 2005, Proceedings of the 30$^{\rm{th}}$ 
Int.~Cosmic Ray Conf.~2007, 2, 633

\bibitem[Hinton \& Aharonian (2007)]{Hinton2007}Hinton, J.A., \& 
Aharonian, F.A. 2007, ApJ, 657, 302

\bibitem[Hofmann et al.~(1999)]{Hofmann1999}Hofmann, W., et al. 1999, 
APh, 12, 135

\bibitem[Holder et al.~(2008)]{VERITAS}Holder, J., Acciari, V.A., Aliu, 
E., et al. 2008, in AIP Conf. Proc. 1085, Proc. 4th Int. Meeting on High 
Energy Gamma-Ray Astronomy, ed. F. A. Aharonian, W. Hofmann, \& F. 
Rieger (Melville, NY: AIP), 657

\bibitem[Jungman et al.~(1996)]{Neutralino}Jungman, G., Kamionkowski, 
M., \& Griest, K. 1996, PhR, 267, 195

\bibitem[Kosack et al.~(2004)]{Whipple_GC}Kosack, K.P., Badran, H.M., 
Bond, I.H., et al. 2004, ApJ, 608, 97

\bibitem[Kosack et al.~(2005)]{Kosack_Thesis}Kosack, K.P. 2005, PhD 
thesis, Washington University in St.Louis

\bibitem[Kusunose \& Takahara (2012)]{Kusunose2012}Kusunose, M., \& 
Takahara, F. 2012, ApJ, 748, 34

\bibitem[Linden et al.~(2012)]{Linden2012}Linden, T., Lovegrove, E., \& 
Profumo, S. 2012, ApJ, 753, 41

\bibitem[Livio \& Silk (2014)]{HighMass_DM}Livio, M., Silk, J. 2014, 
Nature, 507, 29

\bibitem[Muno et al.~(2005)]{Muno2005} Muno, M.P., Lu, J.R., Baganoff, 
F.K., et al. 2005, ApJ, 633, 228

\bibitem[Neilsen et al.~(2013)]{Chandra2012}Neilsen, J., Nowak, M.A., 
Gammie, C., et al. 2013, Proceedings of IAU Symposium No. 303, The 
Galactic Center: Feeding and Feedback in a Normal Galactic Nucleus, see 
arXiv1311.6818

\bibitem[Nolan et al.~(2012)]{Fermi_SecondCatalog}Nolan, P.L., Abdo, 
A.A., Ackermann, M., et al. 2012, ApJS, 199, 31

\bibitem[Ponti et al.~(2010)]{Fluorescense2} Ponti, G., Terrier, R., 
Goldwurm, A., B{$\acute{\rm{e}}$}langer, G., Trap, G. 2010, ApJ, 714, 
732

\bibitem[Porquet et al.~(2005)]{Porquet2005}Porquet, D., Grosso, N., 
Burwitz, V., et al. 2005, A\&A, 430, 9

\bibitem[Saitoh et al.~(2014)]{G2_Simulations3} Saitoh, T.R., Makino, 
J., Asaki, Y., et al. 2014, PASJ, 66, 1

\bibitem[Sakano et al.~(2005)]{Sakano2005}Sakano, M., Warwick, R.S., 
Decourchelle, A., Wang, Q.D. 2005, MNRAS, 357, 1211

\bibitem[Sunyaev et al.~(1993)]{Fluorescense3} Sunyaev, R.A., 
Markevitch, M., Pavlinsky, M. 1993, ApJ, 407, 606

\bibitem[Su et al.~(2010)]{FermiBubbles}Su, M., Slatyer, T.R., 
Finkbeiner, D.P. 2010, ApJ, 724, 1044

\bibitem[Supanitsky (2014)]{Supanitsky2014}Supanitsky, A.D. 2014, PhRvD, 
89, 023501

\bibitem[Terrier et al.~(2010)]{Fluorescense1} Terrier, R., Ponti, G., 
B{$\acute{\rm{e}}$}langer, G., et al. 2010, ApJ, 719, 143

\bibitem[Tsuchiya et al.~(2004)]{CANGAROO_GC}Tsuchiya, K., Enomoto, R., 
Ksenofontov, L.T., et al. 2004, ApJ, 606, L115

\bibitem[Vasileiou et al.~(2011)]{Fermi_GC_Burst}Vasileiou, V., Chiang, 
J., Omodei, N., Piron, F., Vianello, G., McEnery, J. 2011, Astronomer's 
telegram \#3162

\bibitem[Viana \& Moulin (2013)]{HESS_UpdatedSpectrum}Viana, A., \& 
Moulin, E. 2013, Proceedings of the 33$^{\rm{rd}}$ Int.~Cosmic Ray 
Conf.~2013, see 
http://143.107.180.38/indico/contributionDisplay.py?contribId=901\&sessionId=3\&confId=0

\bibitem[Wang et al.~(2006)]{GC_Plerion}Wang, Q.D., Lu, F.J., Gotthelf, 
E.V., et al. 2006, MNRAS, 367, 937

\bibitem[Wijnands et al.~(2006)]{Wijnands2006} Wijnands, R., Zand, 
J.J.M., Rupen, M., et al. 2006, A\&A, 449 , 1117

\bibitem[Yang et al.~(2013)]{FermiBubbles_AGN} Yang, H.-Y.K., 
Ruszkowski, M., Zweibel, E. 2013, MNRAS, 436, 2734

\bibitem[Yoshikoshi et al.~(2009)]{CangarooSystematic}Yoshikoshi, T., et 
al. 2009, ApJ, 702, 631

\bibitem[Yusef-Zadeh et al.~(2013)]{Fermi_Contours}Yusef-Zadeh, F., et 
al. 2013, ApJ, 762, 33

\end{thebibliography}
\end{document}